\documentclass[10pt]{article}
\usepackage[OE]{express}

\begin{document}

\title{Polarization-dependent electromagnetic responses of ultrathin and highly flexible asymmetric terahertz metasurfaces}
 
\author{Joshua A. Burrow\authormark{1,4}, Riad Yahiaoui\authormark{2}, Andrew Sarangan\authormark{1}, Imad Agha\authormark{1,3}, Jay Mathews\authormark{3} and Thomas A. Searles\authormark{2,5}}

\address{\authormark{1}Electro-Optics Department, University of Dayton, 300 College Park Avenue, Dayton, OH, 45469, USA\\
\authormark{2}Department of Physics $\&$ Astronomy, Howard University, 2355 6th St. NW, Washington, DC, 20059, USA\\
\authormark{3}Physics Department, University of Dayton, 300 College Park Avenue, Dayton, OH, 45469, USA\\
\authormark{4}burrowj2@udayton.edu\\
\authormark{5}thomas.searles@howard.edu}

\begin{abstract}
We report the polarization-dependent electromagnetic response from a series of novel terahertz (THz) metasurfaces where asymmetry is introduced through the displacement of two adjacent metallic arms separated by a distance $\delta$. For all polarization states, the symmetric metasurface exhibits a low quality (Q) factor fundamental dipole mode. By breaking the symmetry, we experimentally observe a secondary dipole-like mode with a Q factor nearly $9\times$ higher than the fundamental resonance. As $\delta$ increases, the fundamental dipole mode $f_{1}$ redshifts and the secondary mode $f_{2}$ blueshifts creating a highly transmissive spectral window. Polarization-dependent measurements reveal a full suppression of  $f_{2}$ for all asymmetries at $\theta \geq 60^\circ$. Furthermore, at  $\delta \geq 60 \text{ }\mu\text{m}$, we observe a polarization selective electromagnetic induced transparency (EIT) for the fundamental mode. This work paves the way for applications in filtering, sensing and slow-light devices common to other high Q factor THz metasurfaces with EIT-like response. 
\end{abstract}



\section{Introduction}
Symmetry plays an important role in many facets of modern society including art, architecture, fashion and mathematics.  
Similarly, breaking symmetry can also lead to important implications with respect to the physical properties of materials. Recently, physicists have demonstrated the importance of asymmetry in the defects of crystals \cite{Medlin17}, the transport properties of strained topological insulators \cite{Jost17} and the electromagnetic response of metamaterials \cite{Fedotov07,Singh10,Jansen11,Singh11,Al-Niab12,Singh14,Born14,Manjappa15,Yang17,Yin13,Zhao17}. Specifically for metamaterials, asymmetric unit cells result in Fano-like high 
Q factor modes desirable for applications in communications and sensing. Several methods have been explored for the achievement of high 
Q factor resonances with respect to planar metamaterials such as capacitative gap translation\cite{Cong15,Singh10,Manjappa15,Srivastava15}, toroidial designs\cite{Gupta16,Gupta17,Cong17}, varying the periodicity \cite{Xu16,Yang16}, and mirrored asymmetric macro-cells\cite{Al-Niab12,Yang16}. 


For asymmetric THz planar metamaterials, Singh \textit{et al.} showed an increase in the Q factor of a quadropole mode when the incident  \textbf{E}-field is horizontally oriented with respect to the capacitive gap versus when the incident  \textbf{E}-field is vertically polarized \cite{Singh10}. In subsequent studies for asymmetric metasurfaces, the polarization dependence for normally incident parallel and perpendicular  \textbf{E}-fields have shown transmission and frequency modulation along with excitation of new high Q resonances \cite{Singh11,Al-Niab12}.  However, to our knowledge, there has not been a study with respect to multiple polarized orientations of the incoming THz wave for asymmetric metasurfaces. Polarization-dependent measurements on A-shaped resonators revealed that transmission windows arose for completely different physical mechanisms from horizontally  and perpendicularly polarized incident waves \cite{Zhang17}. Specifically for the perpendicular polarization, the transmission window stemmed from the plasmon-induced transparency effect whereas for horizontal polarization the window appeared due to the coupling of a localized LC resonance and a nearby higher order mode.

In this article, the polarization dependence of the THz response for ultrathin and highly flexible asymmetric metasurfaces is explored experimentally and presented with supporting numerical simulations. Due to the extreme mechanical flexibility of the substrate, the investigated structure is very appropriate for non-planar applications. Recently, there have been various efforts devoted to demonstrate flexible metamaterials\cite{Yahiaoui13,Yahiaoui14,Yahiaoui15}. The use of ultrathin and highly flexible substrates has provided an unprecedented route to achieve active tunability in the frequency of metamaterials due to modifications in the profiles and the periodicities of the structures when the substrates are stretched \cite{Tao08, Lee12, Li13, Zhang15}.  Here, asymmetry is introduced in a novel way by shifting the metal braces rather than shifting the position of the capacitive gap. From  this  approach, there is a significant amplitude modulation of a secondary dipole-like resonance as $\delta$ is increased. In addition, this high Q factor mode decreases linearly in Q and increases exponentially in modulation depth as one increases the shift.  Furthermore, there is the presence of a polarization selective EIT seen for the fundamental mode at high asymmetries.    

\section{Materials and Methods}

\begin{figure}[ht!]
\centering\includegraphics[width=\linewidth]{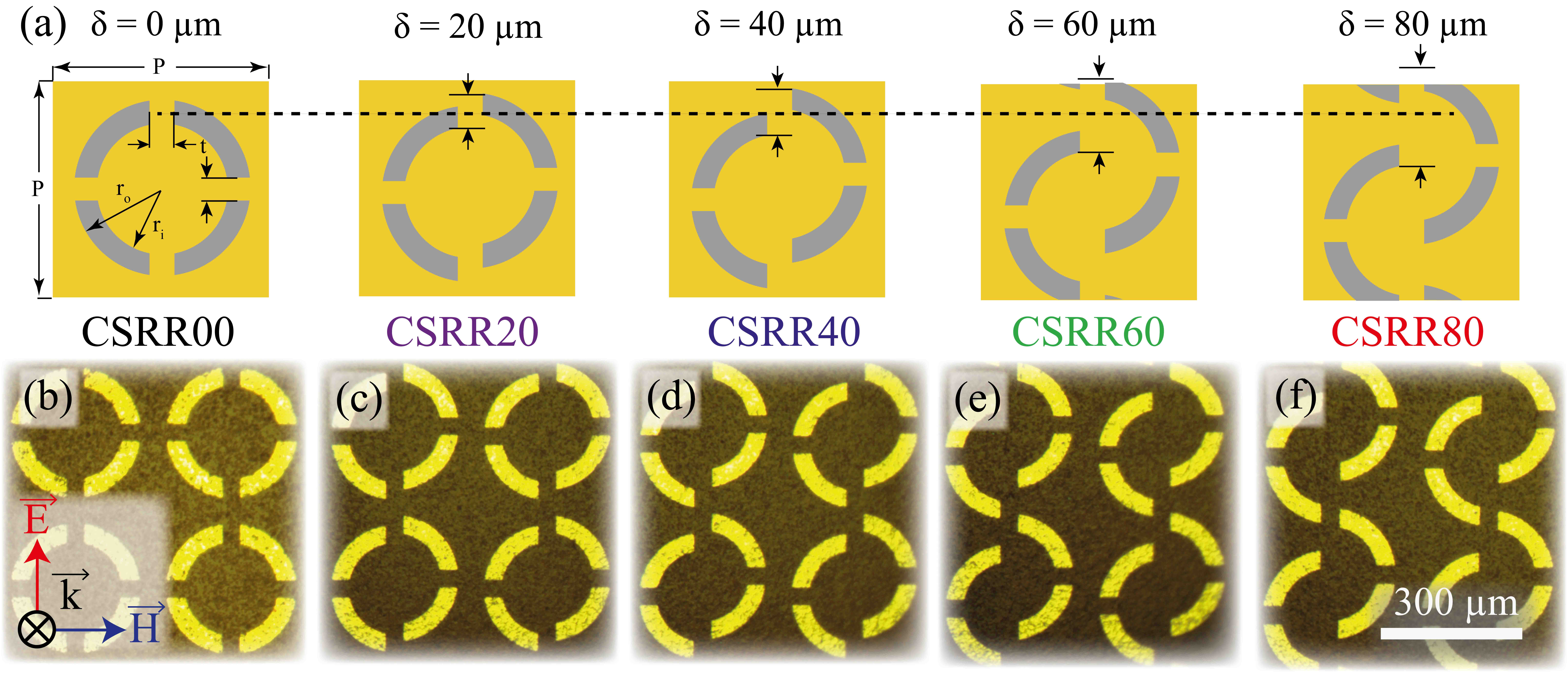}
\caption{(a) Geometrical description of the CSRR unit cell indicating the asymmetry parameter, and microscope images of CSRR (b) $\delta = 0\text{ }\mu\text{m}$, (c) $\delta = 20\text{ }\mu\text{m}$, (d) $\delta = 40\text{ }\mu\text{m}$, (e) $\delta = 60\text{ }\mu\text{m}$, and (f) $\delta = 80\text{ }\mu\text{m}$.}
\label{fig:1}
\end{figure}

The unit cell for the symmetric circle split ring resonator at $\delta = 0\text{ }\mu\text{m}$ (CSRR00) is shown in Fig. \ref{fig:1}(a). It features four capacitative gaps uniformly distributed along a metallic ring with the following geometric parameters: periodicity
$\text{P} = 300\text{ }\mu\text{m}$, outer ring radius $r_{o} = 125\text{ }\mu\text{m}$, inner ring radius $r_{i} = 107.5 \text{ }\mu\text{m}$ and square gap length $t = 35 \text{ }\mu\text{m}$. The unit cell is then repeated periodically to form a metasurface of 100 nm thick Ag structures deposited using electron beam evaporation and patterned using optical lithography on an ultra-flexible polyimide substrate of $50.8\text{ } \mu\text{m}$ thickness. Such periodic structures do not diffract normally incident electromagnetic radiation for frequencies lower than 1 THz. Microscope images of the resulting devices are depicted in Fig. \ref{fig:1}(b)-(f) for $\delta = 0,\text{ } 20,\text{ } 40,\text{ } 60, \text{ and } 80\text{ }\mu\text{m}$, where the asymmetry parameter $\delta$ indicates the translational offset between two adjacent metallic arms. The total active surface area of each of the fabricated devices was 1 cm $\times$ 1 cm.

The planar metamaterial devices were characterized using a continuous-wave (CW) THz spectrometer  (Teraview CW Spectra 400) that emits linearly polarized collimated THz radiation. In this setup, spectroscopy is achieved by incrementally varying the difference frequency of two near-IR diode lasers via a temperature tuning technique that is smooth and mode hop free for a spectral resolution of 100 MHz. The nominal spectral resolution is then governed by the precision of the laser frequency control and not by a mechanical delay stage as found in a conventional THz time-domain spectroscopy (TDS) setup. With a spectral resolution higher than THz TDS, we were able to resolve extremely narrow spectral features. The transmission from each sample was determined as $T(f) = P_{MM}(f)/P_{sub}(f)$, where $P_{MM}(f)$ and $P_{sub}(f)$ are the filtered THz power spectra of the planar metamaterial and polyimide substrate respectively. To investigate the polarization dependence of the symmetric and asymmetric devices, each sample was rotated by an angle $\theta$ to mimic various incident polarization states in which $\theta$ represents the angle from the vertical polarization. 

To compliment the experimental results, numerical calculations were carried out using CST Microwave Studio \cite{CST}, a finite difference time domain based technique used to model the governing equations of electrodynamics. In these calculations, the primitive cells of the designed metasurfaces were illuminated by a normally incident plane wave with an appropriate polarization of the electric and magnetic fields, as depicted in the inset of Fig. \ref{fig:1}(b). Periodic boundary conditions were imposed in the numerical model to mimic a 2D infinite structure.  The material parameters used for the lossy Ag metal layer and ultra-flexible dielectric substrate were $\sigma = 6.1\times10^{7} \text{ S/m}$ and $\epsilon = 3.3 + i0.03$, respectively.    

\section{Results and Discussions}
\begin{figure}[ht!]
\centering\includegraphics[width=\linewidth]{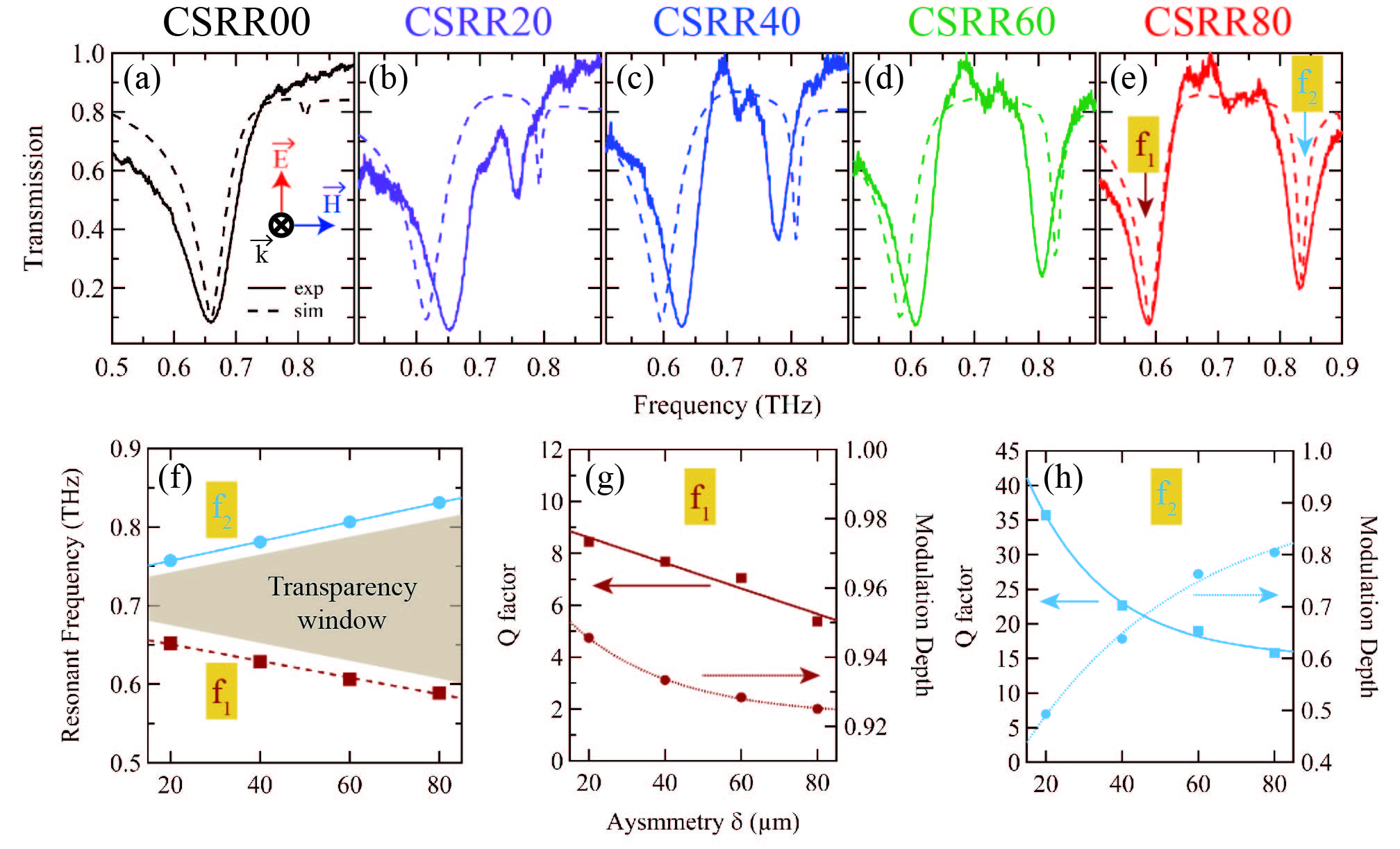}
\caption{Measured (solid) and simulated (dashed) transmission spectra under $E_{y}$ polarization for (a) CSRR00, (b) CSRR20, (c) CSRR40, (d) CSRR60, and (e) CSRR80. Experimentally measured resonant frequency (f), (g) Q factor and modulation depth of $f_{1}$ and (h) Q factor and modulation depth of $f_{2}$ as a function of $\delta$.}
\label{fig:2}
\end{figure}

Fig. \ref{fig:2}(a)-(e) compares the measured (solid) and simulated (dashed) transmission spectra under vertical \textbf{E}-field polarization for CSRR00, CSRR20, CSRR40, CSRR60, and CSRR80. For the symmetric geometry (CSRR00), a fundamental resonant frequency $f_{1} \sim 0.65\text{ THz }$ was observed with an experimental quality factor of $Q = 4.13$.  Due to the four-fold rotational symmetry imposed on the design, the performance of the device was insensitive to the polarization of the incident wave. After introducing asymmetry ($\delta \geq 20 \mu\text{m}$), $f_{1}$ redshifts with increasing asymmetry and a second mode $f_{2}$ is excited near 0.76 THz. For $f_{2}$, there was a blueshift in frequency and the transmission exponentially increased with increasing $\delta$. The latter result infers an enhancement in  modulation depth for $f_{2}$ that is not seen for $f_{1}$.  In general, the experimental results are well supported by numerical simulations.  However, the numerical results indicate the existence of $f_{2}$ in the symmetric case, which is weakly excited and almost imperceptible in simulation and not observed experimentally.  Other slight discrepancies in the results of simulations and measurements as seen in Fig. 2 can be attributed to small variations in the unit cells of each metasurface after fabrication and dispersions being unaccounted for during simulations.

A Fano fit was applied to each resonance with the form:
\begin{equation}
F(f) = A\frac{[\Lambda + (f - f_{0})/\Delta f]^{2}}{1 + (f - f_{0})^{2}}
\end{equation}
where $f_{0}$ is the resonant frequency, $\Lambda$ yields the lineshape asymmetry parameter and $\Delta f$ is the full width half maximum (FWHM) of the mode \cite{Zhang11}. With the extracted Fano lineshape parameters from Eqn. 1, the Q factor and modulation depth for $f_{1}$ and $f_{2}$ are calculated and compared to one another with respect to the asymmetry parameter $\delta$. The Q factor of each resonance is defined as $Q = f_{i}/\Delta f \text{ for } i \in \{1,2\}$ where $f_{i}$ is the resonant frequency and $\Delta f$ is the FWHM.  By tailoring the asymmetry in the structure, the frequency shift, modulation depth and Q factor of both $f_{1}$ and $f_{2}$ are tuned and each attribute is plotted in Fig. \ref{fig:2}(f), (g) and (h), respectively.  Both $f_{1}$ and $f_{2}$ shift linearly with increasing asymmetry at a rate of -1.058 GHz/$\mu$m for $f_{1}$ and  1.236 GHz/$\mu$m for  $f_{2}$ as illustrated in Fig. \ref{fig:2}(f). Furthermore, there is an increasing flattop ultra-broadband transmission window as the asymmetry increases indicated by the shaded area in Fig. \ref{fig:2}(f). At $\delta = 80 \mu$m, the transmission bandwidth is maximal at $\sim$ 200 GHz. This result shows great promise for these asymmetric metasurfaces for applications as frequency selective devices, or THz bandpass filters.
 
With respect to Q factor, at $\delta = 20 \mu$m the Q factor for $f_{1}$ is 8.4 and decreases to 5.3 at 80 $\mu$m. These values are on the order of fundamental modes (LC, dipole, etc.) for symmetric metasurfaces \cite{Singh10,Xu16}.  The Q factor of $f_{2}$ is 35.7 at 20 $\mu$m and decreases exponentially to 15.8 at 80 $\mu$m.  As a comparison, the Q factor observed for the asymmetric case at 20 $\mu$m is one of highest experimental Q factors reported and more than 8$\times$ greater than that of $f_{1}$. For THz metasurfaces, high experimental Q factors have been reported for toroidal (Q = 42.5)\cite{Gupta16} and low asymmetric gap translations (Q = 49)\cite{Srivastava15} while other high Q factor modes introduced by asymmetry are on the order of 20-30 \cite{Fedotov07,Al-Niab12,Al-Naib15}. To evaluate how the Q factors evolve with increasing asymmetry, a linear fit was applied to the Q factor relation for $f_{1}$ revealing a $-0.049\mu\text{m}^{-1}$ slope. Similarly, for the Q factor of $f_{2}$, an exponential decay  was observed and the fit equation was given by $Q = 15.05 + 52.60\exp-0.047\delta$. Decreases in Q factor with increasing degrees of asymmetry have previously been reported \cite{Fedotov07,Singh11,Jansen11} and attributed to increases in radiation loss from an enlarged variation in the surface charge distributions between two excited split-ring resonators and a resultant suppression of the dipole moment in the far field \cite{Yang17}.  To further analyze the results, the modulation depth was defined as $MD = A - \text{min}(T)$ where $A$ is the maximum value of the asymmetric lineshaped transmission dip nearest to the resonant frequency. A proportional relationship between modulation depth and asymmetry was previously reported for trapped and octopolar modes\cite{Yang17,Al-Naib15}.  For $f_{1}$, there is a subtle exponential decrease from 0.95 at 20 $\mu$m to 0.93 at 80 $\mu$m indicating that the coupling strength of $f_{1}$ is nearly unaffected by the asymmetric modification. However, for $f_{2}$, there was enhancement of the modulation depth up to 40\% with increasing $\delta$. 

\begin{figure}[ht!]
\centering\includegraphics[width=\linewidth]{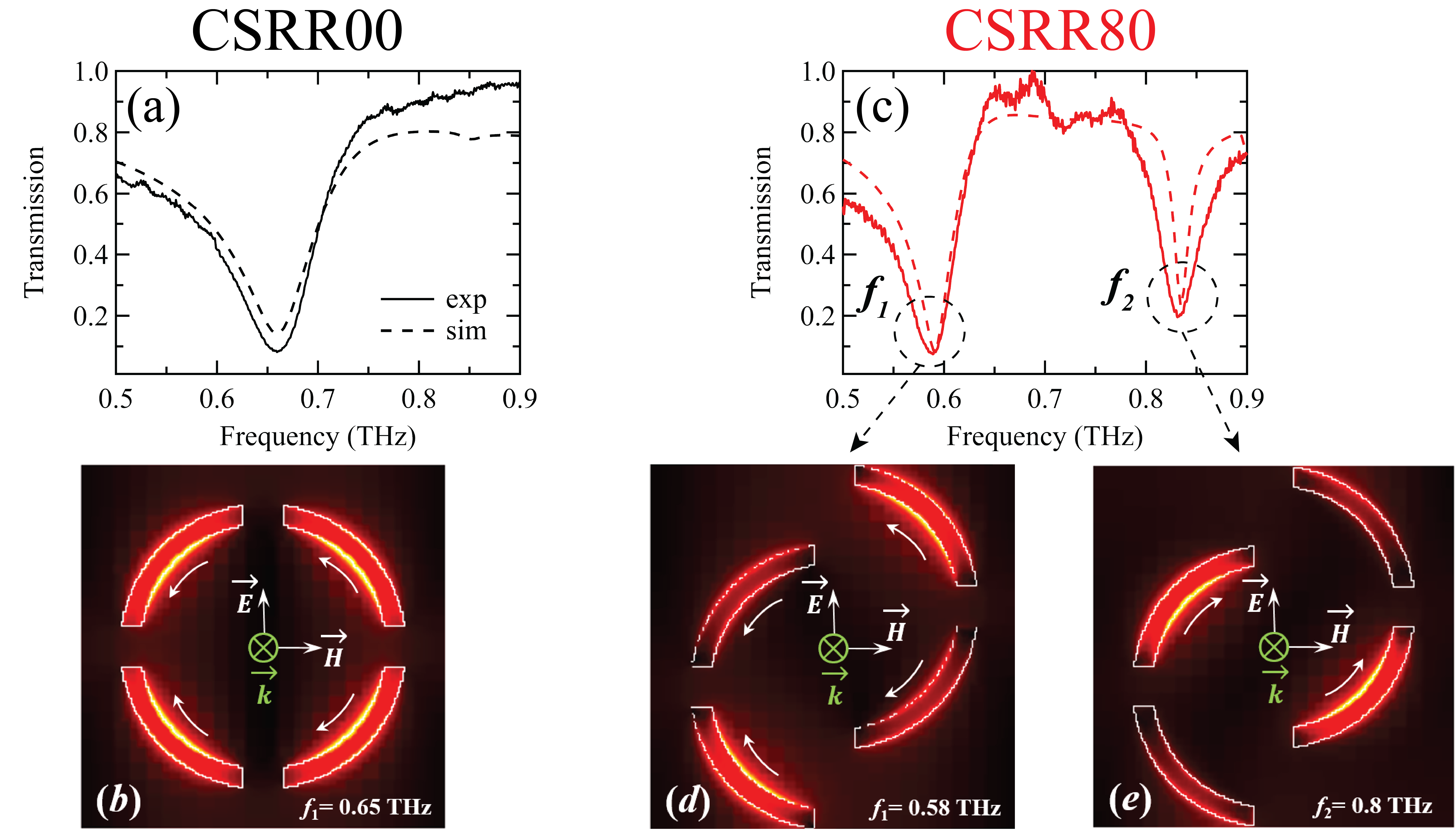}
\caption{CSRR00 (a) experimental and simulation transmission spectra and (b) surface current (white arrows) and magnetic field distribution $\sim 0.65$ THz. CSRR80 (c) experimental and simulation transmission spectra, surface currents (white arrows) and magnetic field distributions at (d)$f_{1}\sim0.58$ THz and (e) $f_{2}\sim 0.8$ THz.}
\label{fig:3}
\end{figure}

To gain understanding of the physical mechanism for the electromagnetic responses, the surface currents and magnetic field distributions for each metasurface were simulated and results for the CSRR00 and CSRR80 for vertical \textbf{E}-field polarization are shown in Fig.~\ref{fig:3}(b), (d), and (e).  For the CSRR00 device, the surface current \textbf{J} of $f_{1} \sim 0.65$ THz portrays a strong dipolar nature with $\textbf{J}$ converging to the left gap.  This is similar to the dipole resonance seen in other two gap CSRRs \cite{Fedotov07,Born14}.  For the CSRR80 device near $f_{2} \sim 0.8$ THz, the simulated surface current~\textbf{J} and magnetic field |\textbf{H}| amplitude indicate a dipole oscillation in the upper left and lower right arms of the CSRR where the surface current converges to the opposite gap from $f_{1}$.  Intuitively, the redshift of the fundamental mode $f_{1}$ could be attributed to the increase in the surface current path of the metasurface, while $f_{2}$ is blueshifted because of the decrease in the current path.  These frequency shifts of opposite signs suggest that $f_{2}$ originated from the structural asymmetry of each meta-atom rather than from mimicking its fundamental mode. Indeed, if $f_{2}$ originated from other higher order resonances, $f_{2}$ must be redshifted as much as $f_{1}$ is redshifted because all the higher order frequencies depend on the wavelengths of their fundamental frequencies. However, in our structure, upon increasing the asymmetry ($\delta \geq 20\text{ } \mu\text{m}$) $f_{1}$ is redshifted while $f_{2}$ is blueshifted, thus giving rise to a broadband transparency window [see Fig. 3(c)].  Moreover, $f_{2}$ vanished for the CSRR00 when the current paths became perfectly symmetrical. Similar behavior has previously been reported in anisotropic THz metamaterials \cite{Jung14}.

\begin{figure}[ht!]
\centering\includegraphics[width=0.9\linewidth]{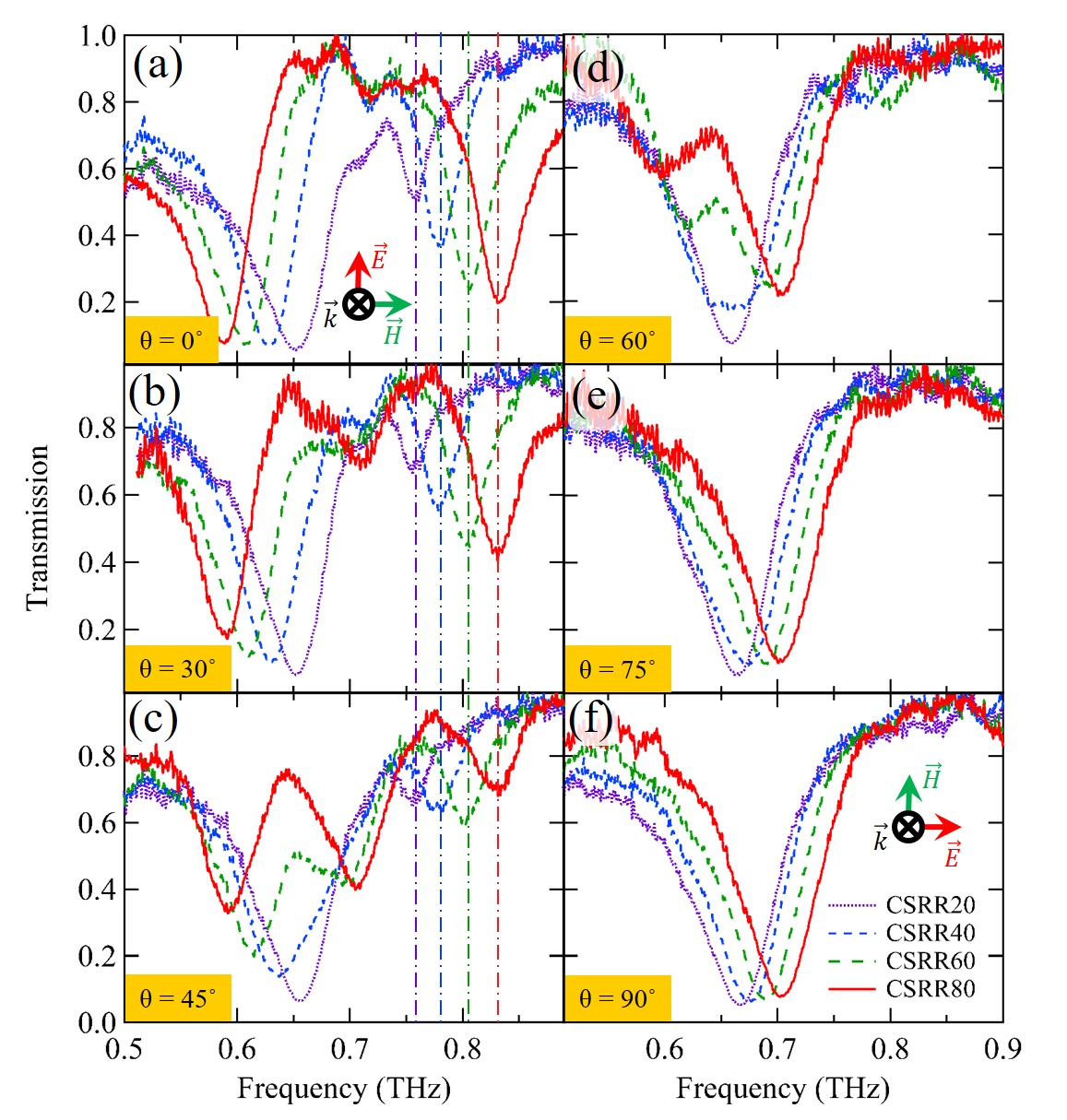}
\caption{Evolution of the measured transmission spectra for different values of the shift $\delta$ = 20, 40, 60 $\text{and } 80\text{ }\mu\text{m}$, respectively and under different polar angles $\theta$ varying from $0^{\circ}$ to $90^{\circ}$.}
\label{fig:4}
\end{figure}

To further understand the nature of the resonances, the effect of polarization was measured by rotating each device to mimic various linear polarization states.  Fig. \ref{fig:4}(a) shows the transmission spectra for $\delta = 20\text{ (purple)}, 40\text{ (blue)}, 60\text{ (green)} \text{ and }80\text{ } \mu$m (red) under a vertically polarized THz wave and the \textbf{E} is parallel to the direction of asymmetric shift.  Fig. \ref{fig:4}(b) - (f) depicts the experimental transmittance as the device was rotated counter clockwise  by $\theta$ until the $\textbf{E}$ field was orthogonal to shift direction.  For the cases when $\delta = 20 \text{ and } 40 \text{ }\mu\text{m}$, the resonant frequency and minimum transmission amplitude of $f_{1}$ oscillates sinusoidally with respect to polar angle $\theta$ and can easily be understood by Malus' Law. However, for $\delta \geq 60 \text{ }\mu\text{m}$ under the cross polarization ($\theta = 45^{\circ}$) excitation state, a mode splitting of $f_{1}$ was observed.  This polarization selective splitting observed for high asymmetries is indicative of electromagnetic induced transparency (EIT) previously seen for  A-shaped THz metasurfaces \cite{Zhang17} and other plasmonic systems \cite{Yahiaoui17,Amin15}. The change in coupling strength of the left and right modes for $f_{1}$ continues for $\theta = 60^{\circ}$and then returns to one resonance above $\theta = 75^{\circ}$. Mode $f_{2}$ solely exhibits transmittance amplitude modulation and is fully suppressed when the $\textbf{E}$-field is orthogonal to the direction of translation. The vertical dotted lines indicate the frequency invariance (i.e. the frequency is independent of the polarization) of $f_{2}$ for all asymmetric cases where $\theta \leq 45^{\circ}$. The resonant frequencies of $f_{2}$ are 0.76, 0.78, 0.81, and 0.83 THz for CSRR20, CSRR40, CSRR60, and CSRR80. It is well known that frequency invariance with respect to the polarization insensitivity of high Q factor modes can be advantageous to metasurface sensing applications \cite{Wang16}.

As previously stated, when $\theta = 45^{\circ}$ and upon increasing the asymmetry, the spectral response of the metasurface has a similar signature to the EIT effect, where a transmission peak appears at about 0.657 THz with an amplitude of $57\%$ between two resonance dips at around 0.587 THz and 0.7 THz, respectively [see Fig. 4(c)].  ~~Figure 5(a)-(c) shows the simulated (dashed lines) and measured (solid lines) transmission spectra for different values of the asymmetry $\delta = 0, 60$ and $80~\mu\text{m}$, respectively and under a fixed polar angle of $45^{\circ}$.  Although there are minor differences in amplitude and bandwidth of the resonances (probably due to microscopic imperfections induced during the fabrication process), the transmission spectra obtained from numerical simulations confirm the trends obtained from measured data.  To gain a deeper insight into the physics of the EIT-like effect, we have plotted in Figs. 5(e)-(g) the magnetic field distributions at $f_{l}$, $f_{t}$ and $f_{h}$, respectively.  Here, we define the lower resonance frequency, the transmission peak, and the higher resonance frequency as $f_{l}$, $f_{t}$ and $f_{h}$.  To better guide the eye, the directions of the THz induced surface currents are indicated by white arrows in Figs. 5(e)-(g).  For the lower frequency dip at $f_{l}$, the CSRR is highly radiative. It exhibits parallel surface current distributions on the right and left arms and on the upper and lower arms as well.  This surface current distribution possesses a strong field confinement suggesting direct dipole coupling with the external incident radiation wave. 

\begin{figure}[ht!]
\centering\includegraphics[width=\linewidth]{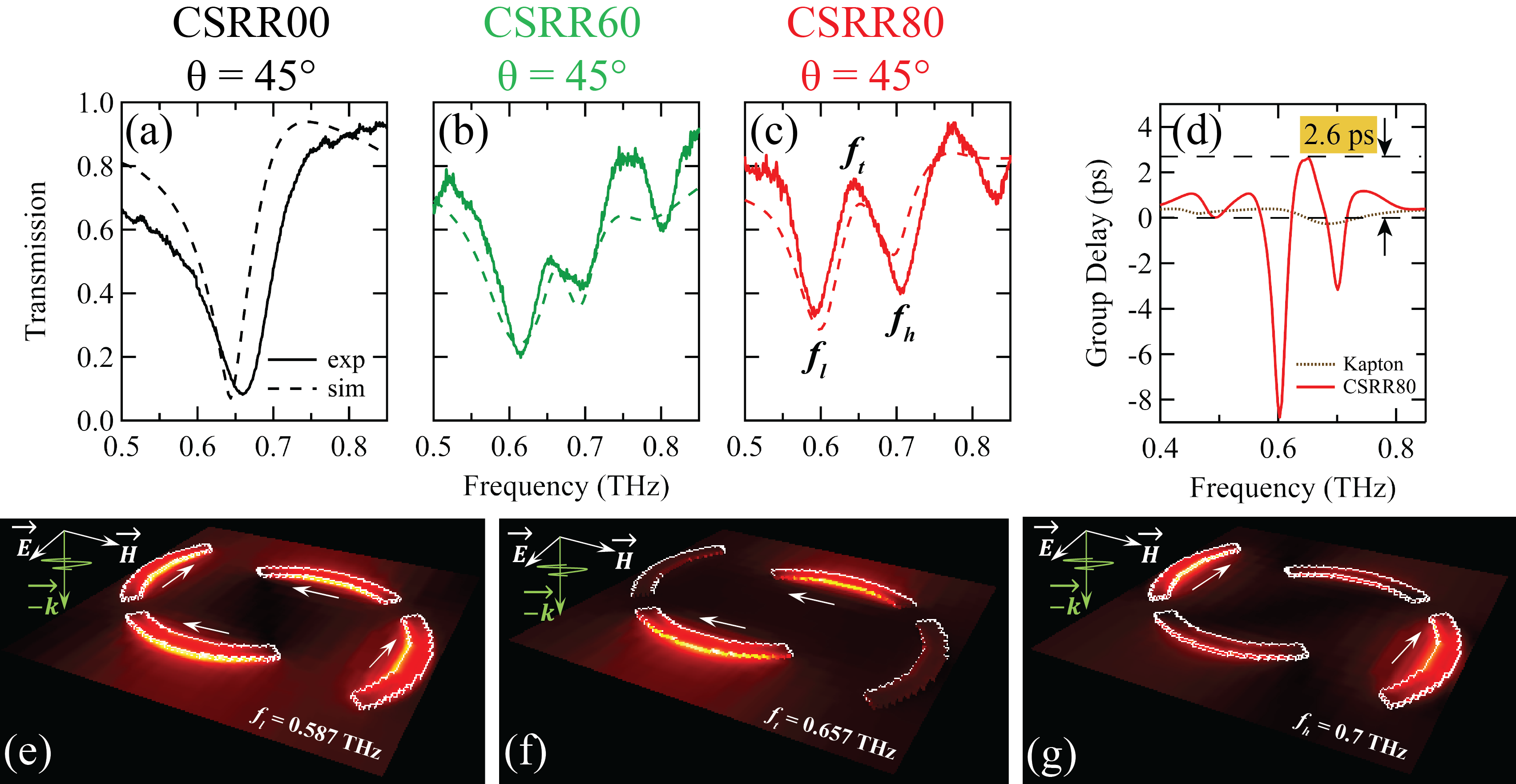}
\caption{Measured (solid) and simulated (dashed) transmission spectra excited by cross polarized THz radiation for (a) CSRR00, (b) CSRR60, and (c) CSRR80 devices. (d) Simulated group delays associated with the CSRR80 (red solid line) that exhibit EIT at 0.657 THz and the reference kapton film (gold dashed line), respectively. (e)-(g) Magnetic field distributions at $f_{l}$, $f_{t}$ and $f_{h}$, respectively for the CSRR80. The corresponding induced surface currents are indicated by white arrows.}
\label{fig:5}
\end{figure}

Regarding the transmission peak at $f_{t}$, it is found that the surface currents in the right and left arms are strongly suppressed and mainly flow along the upper and lower arms, as shown in Fig. 5(f).  The corresponding magnetic field mainly concentrates in the upper and lower arms as well.  In this case, the CSRRs couple weakly with the incident THz radiation.  Therefore, it can be deduced that the transparency window appears due to the suppression of the dipole response of the right and left arms of the CSRRs.  For the higher frequency dip at $f_{h}$, the surface currents flow along the right and left arms, where the corresponding magnetic field is mainly localized [see Fig. 5(g)].  Thus, it can be deduced from the current and magnetic field distributions that dipole oscillations are excited by the incident wave for $f_{h}$. 

Although the experimental portion of this investigation was carried out using a CW source, it is worth noting the group delay response for pulsed THz slow light applications.  The group delay $\tau_{g}$, retrieved from the simulated transmission spectrum of CSRR80, is represented in Fig. 5(d).  The group delay of the kapton substrate is also plotted in Fig. 5(d) for comparison.  The group delay was calculated using $\tau_{g}=-d\phi(\omega)/d\omega$, where $\phi(\omega)$ is the transmission phase and $\omega=2\pi f$ with $f$ representing the frequency.  One can observe that at frequencies away from the resonances, the group delay of THz radiation tends to be equal to that of kapton film.  At the resonance frequencies of 0.587 THz and 0.7 THz the metamaterial demonstrates negative group delay. In the vicinity of the transparency peak, large positive group delays are obtained, indicating potential use in slow light applications.  For example, at around 0.657 THz where 57\% of EIT transmission is achieved, the THz radiation experiences a delay of about 2.6 ps, which is equivalent to the time delay of a 780 $\mu\text{m}$ distance of free space propagation.

\section{Conclusions}

In summary, we have experimentally and numerically characterized the polarization-dependent electromagnetic responses of ultra-flexible asymmetric THz metasurfaces and demonstrated a high experimental Q factor ($\sim$35.7) resonance in the transmission spectrum.   The modulation depth of the asymmetric mode was increased by displacing adjacent metallic arms and also by modifying the interaction between the metasurface and the incident linear polarization state of incident THz radiation.  The induced Fano shaped resonance was investigated through simulations, polarization dependence and asymmetry, revealing the dipole like and frequency invariant nature of the mode.  The asymmetric metasurfaces studied here show great promise for low cost devices based platforms that include THz bandpass filters, biological sensors and slow-light devices. 

\section*{Funding}
National Science Foundation (NSF) (ECCS-1541959, ECCS-1710273, ECCS-1709200); NASA Ohio Space Grant (NNX15AL50H); Air Force Office of Scientific Research (FA9550-16-1-0346). 

\section*{Acknowledgments}
We would like to acknowledge support from Air Force Research Laboratory Minority Leadership Program. This work was performed in part at the Harvard Center for Nanoscale Systems (CNS), a member of the National Nanotechnology Coordinated Infrastructure Network (NNCI). T. A. S. acknowledges support from the CNS Scholars Program. 


\begin{thebibliography}{99}



\bibitem{Medlin17} D. L. Medlin, K. Hattar, J.A. Zimmerman, F. Abdeljawad, and S. M. Foiles, ``Defect character at grain boundary facet junctions: Analysis of an asymmetric $\Sigma = 5$ grain boundary in Fe,'' In Acta Mater. {\bfseries 124}, 383--396 (2017).


\bibitem{Jost17} A. Jost, M. Bendias, J. B\"{o}ttcher, E. Hankiewicz, C. Brune, H. Buhmann, L. W. Molenkamp, J. C. Maan, U. Zeitler, N. Hussey, and S. Wiedmann, ``Electron-hole asymmetry of the topological surface states in strained HgTe,'' Proc. Natl. Acad. Sci. U.S.A. {\bfseries 114}(13), 3381--3386 (2017).


\bibitem{Fedotov07} V. A. Fedotov, M. Rose, S. L. Prosvirnin, N. Papasimakis, and N. I. Zheludev, ``Sharp trapped mode resonances in planar metamaterials with a broken structural symmetry,'' Phys. Rev. Lett. {\bfseries 99}(14), 147401 (2007).

\bibitem{Singh10} R. Singh, I. Al-Naib, M. Koch, and W. Zhang, ``Asymmetric planar terahertz metamaterials,'' \opex {\bfseries 18}(12), 13044--13050 (2010). 

\bibitem{Singh11} R. Singh, I. Al-Naib, M. Koch, and W. Zhang, ``Sharp Fano resonances in THz metamaterials,'' \opex  {\bfseries 19}(7), 6312--6319 (2011). 

\bibitem{Al-Niab12} I. Al-Niab, R. Singh, C. Rockstuhl, F. Lederer, S. Delprat, D. Rocheleau, M. Chaker, T. Ozaki, and R. Morandotti, ``Excitation of a high-Q subradiant resonance mode in mirrored single-gap asymmetric split ring resonator terahertz metamaterials,'' Appl. Phys. Lett. {\bfseries 101}(7), 071108 (2012). 

\bibitem{Jansen11} C. Jansen, I. Al-Naib, N. Born, and M. Koch, ``Terahertz metasurfaces with high q-factors,'' Appl. Phys. Lett., {\bfseries 98}(5), 051109 (2011). 

\bibitem{Singh14} R. Singh, W. Cao, I. Al-Naib, L. Cong, W. Withayachumnankul, and W. Zhang, ``Ultrasensative terahertz sensing with high Q-factor Fano resonances in metasurfaces,'' Appl. Phys. Lett. {\bfseries 105}(17), 171101 (2014). 

\bibitem{Born14} N. Born, I. Al-Naib, M. Scheller, C. Jansen, J. V. Moloney, and M. Koch, ``Trapped eigen modes in terahertz asymmetric metamolecules,'' 2014 39th International Conference on Infrared, Millimeter, and Terahertz waves (IRMMW-THz), Tucson, AZ, 1-2 (2014).

\bibitem{Manjappa15} M. Manjappa, S. Y. Chiam, L. Cong, A. A. Bettiol, W. Zhang, and R. Singh, ``Tailoring the slow light behavior in terahertz metasurfaces,'' Appl. Phys. Lett. {\bfseries 106}(18), 181101 (2015). 

\bibitem{Yang17} S. Yang, C. Tang, Z. Liu, B. Wang, C. Wang, J. Li, L. Wang, and C. Gu, ``Simultaneous excitation of extremely high-Q-factor trapped and octopolar modes in terahertz metamaterials,'' \opex {\bfseries 25}(14), 15938--15946 (2017). 

\bibitem{Yin13}X. Yin, T. Feng, S. Yip, Z. Liang, A. Hui, J. C. Ho, and J. Li, ``Tailoring electromagnetically induced transparency for terahertz metamaterials: From diatomic to triatomic structural molecules,'' Appl. Phys. Lett. {\bfseries 103}(2), 021115 (2013). 

\bibitem{Zhao17}Z. Zhao, X. Zheng, W. Peng, H. Zhao, J. Zhang, Z. Luo, and W. Shi, ``Localized slow light phenomenon in symmetry broken terahertz metamolecule made of conductively coupled dark resonators,'' Opt. Mat. Express. \textbf{7}(6), 1950--1961 (2017). 

\bibitem{Cong15} L. Cong, M. Manjappa, N. Xu, I. Al-Naib, W. Zhang, and R. Singh, ``Fano Resonances in Terahertz Metasurfaces: A Figure of Merit Optimization,'' Adv. Optical Mater. {\bfseries 3}(11), 1537--1543 (2015).

\bibitem{Srivastava15} Y. K. Srivastava, M. Manjappa, L. Cong, W. Cao, I. Al-Naib, W. Zhang, and R. Singh, ``Ultrahigh-Q Fano Resonances in Terahertz Metasurfaces: Strong Influence of Metallic Conductivity at Extremely Low Asymmetry,'' Adv. Optical Mater. {\bfseries 4}(3). 1--7 (2015). 

\bibitem{Gupta16} M. Gupta, V. Savinov, N. Xu, L. Cong, G. Dayal, S. Wang, W. Zhang, N. I. Zheludev, and R. Singh, ``Sharp Toroidal Resonances in Planar Terahertz Metasurfaces,'' Adv. Mater. {\bfseries 28}(37), 8206--8211 (2016). 


\bibitem{Gupta17} M. Gupta, Y. K. Srivastava, M. Manjappa, and R. Singh, ``Sensing with toroidal metamaterial,'' Appl. Phys. Lett. {\bfseries 110}(12), 121108 (2017). 

\bibitem{Cong17} L. Cong, Y. K. Srivastava, and R. Singh, ``Tailoring the multipoles in THz toroidal metamaterials,'' Appl. Phys. Lett. {\bfseries 111}(8), 081108 (2017).


\bibitem{Xu16} N. Xu, R. Singh, and W. Zhang, ``High-Q lattice mode matched structural resonances in terahertz metasurfaces,'' Appl. Phys. Lett. {\bfseries 109}(2), 021108 (2016). 

\bibitem{Yang16} S. Yang, Z. Liu, X. Xia, Y. E, C. Tang, Y. Wang, and C. Gu, ``Excitation of ultrasharp trapped-mode resonances in mirror-symmetric metamaterials,'' Phys. Rev. B \textbf{93}(23), 235407 (2016). 

\bibitem{Zhang17} N. Zhang, Q. Xu, S. Li, C. Ouyang, X. Zhang, Y. Li, J. Gu, Z. Tian, J. Han, and W. Zhang, ``Polarization-dependent electromagnetic responses in an A-shape metasurface,'' \opex {\bfseries 25}(17) 20689--20697 (2017). 


\bibitem{Yahiaoui13} R. Yahiaoui, J. P. Guillet, F. de Miollis, and P. Mounaix, ``Ultra-flexible multiband terahertz metamaterial absorber for conformal geometry applications,'' \ol \textbf{38}(5), 4988--4990 (2013).

\bibitem{Yahiaoui14} R. Yahiaoui, K. Takano, F. Miyamaru, M. Hangyo, and P. Mounaix,``Terahertz metamolecules deposited on thin flexible polymer: design, fabrication and experimental characterization'' J. Opt. {\bfseries 16}(9), 094014 (2014).

\bibitem{Yahiaoui15} R. Yahiaoui, S. Tan, L. Cong, R. Singh, F. Yan and W. Zhang, ``Multispectral terahertz sensing with highly flexible ultrathin metamaterial absorber,'' J. Appl. Phys. \textbf{118}(8), 083101 (2015). 

\bibitem{Tao08} H. Tao, A. C. Strikwerda, K. Fan, C. M. Bingham, W. J. Padilla, X. Zhang, and R. D. Averitt, ``Terahertz metamaterials on free-standing highly flexible polyimide substrates,'' J. Phys. D {\bfseries 41}(23), 232004 (2008). 


\bibitem{Lee12} S. Lee, S. Kim, T. Kim, Y. Kim, M. Choi, S. H. Lee, J. Kim, and B. Min, ``Reversibly stretchable and tunable terahertz metamaterials with wrinkled layouts," Adv. Mater. {\bfseries 24}(26), 3491--3497 (2012). 


\bibitem{Li13} J. Li, C. M. Shah, W. Withayachumnankul, B. Ung, A. Mitchell, S. Sriram, M. Bhaskaran, S. Chang, and D. Abbott, "Mechanically tunable terahertz metamaterials," Appl. Phys. Lett. {\bfseries 102}(12), 121101 (2013). 

\bibitem{Zhang15} F. Zhang, S. Feng, K. Qiu, Z. Liu, Y. Fan, W. Zhang, Q. Zhao, and J. Zhou, ''Mechanically stretchable and tunable metamaterial absorber," Appl. Phys. Lett. {\bfseries 106}(9), 091907 (2015).

\bibitem{CST} CST Microwave Studio\textregistered, (http://www.cst.com).


\bibitem{Zhang11} J. Zhang, Z. Peng, A. Soni, Y. Zhao, B. Peng, J. Wang, M. S. Dresselhaus, Qihua, and Xiong, ``Raman spectroscopy of few-quintuple layer topological insulator $\text{Bi}_{2}\text{Se}_{3}$ nanoplatelets,'' Nano
Lett., {\bfseries 11}(6), 2407--2414 (2011).

\bibitem{Al-Naib15}I. Al-Naib, Y. Yang, M. M. Dignam, W. Zhang, and R. Singh, ``Ultra-high Q even eigenmode resonances in terahertz metamaterials,'' Appl. Phys. Lett. {\bfseries 106}(1), 011102 (2015). 

\bibitem{Jung14} H. Jung, C. In, H. Choi, and H. Lee, ``Anisotropy modeling of terahertz metamaterials: polarization dependent resonance manipulation by meta-atom cluster,'' Sci. Rep. {\bfseries 4}, 5217 (2014).


\bibitem{Yahiaoui17} R. Yahiaoui, M. Manjappa, Y. K. Srivastava, and R. Singh ``Active control and switching of broadband electromagnetically induced transparency in symmetric metadevices,'' Appl. Phys. Lett. {\bfseries 111}(2), 021101 (2017).

\bibitem{Wang16} S. Wang, L. Xia, H. Mao, X. Jiang, S. Yan, H. Wang, D. Wei, H.-L. Cui, and C. Du, ``Terahertz Biosensing Based on a Polarization-Insensitive Metamaterial,'' IEEE Photon. Tech. Lett. \textbf{9}(28), 986--989 (2016). 















\bibitem{Amin15} M. Amin and A. D. Khan, ``Polarization selective electromagnetic-induced transparency in the disordered plasmonic quasicrystal structure,'' J. Phys. Chem. C {\bfseries 119}(37), 21633-21638 (2015). 

\bibitem{Khan17} A. D. Khan and M. Amin, ``Polarization Selective Multiple Fano Resonances in Coupled T-Shaped Metasurface,'' IEEE Photon. Tech. Lett. {\bfseries 29}(19), 1611--1614 (2017). 


\end{thebibliography}
\end{document}